\documentclass[11pt, letterpaper, preprint]{aastex} 
\usepackage{hyperref}
\usepackage{anysize}
\usepackage{fancyhdr}
\usepackage{tcolorbox}

\newcommand{\kms}{\,km\,s$^{-1}$}
\newcommand{\hi}{\ion{H}{1}}

\marginsize{2.0cm}{2.0cm}{2.0cm}{2.0cm}  

\begin{document}

\large{{\bf High-S/N Quasar Observations with HST/COS: Deep Fields for Spectroscopy }}

\noindent\normalsize{Principal Author: Andrew Fox$^1$ (afox@stsci.edu)\\
Co-authors: Jerry Kriss$^2$, Philipp Richter$^3$, J. Michael Shull$^4$, 
Frances Cashman$^5$, Sapna Mishra$^2$, 
Annelia Anderson$^6$, Nahum Arav$^7$, Ramona Augustin$^8$, 
Kathleen Barger$^9$, Michelle Berg$^9$, Rongmon Bordoloi$^{10}$, 
Sanchayeeta Borthakur$^{11}$, Joseph Burchett$^{12}$, 
Jane Charlton$^{13}$, Hsiao-Wen Chen$^{14}$, 
Christopher Churchill$^{12}$, Ryan Cooke$^{15}$, 
Annalisa de Cia$^{16}$, Gisella de Rosa$^2$,
Romeel Dav\'e$^{17}$, Yakov Faerman$^{18}$, Travis Fischer$^1$, 
David French$^2$, Farhan Hasan$^2$, Svea Hernandez$^1$, 
Cameron Hummels$^{19}$, Sean Johnson$^{20}$, Glenn Kacprzak$^{21}$, 
Vikram Khaire$^{22}$, Doyeon Avery Kim$^2$, Brad Koplitz$^7$, 
Varsha Kulkarni$^{23}$, Nicolas Lehner$^{24}$,
Matilde Mingozzi$^1$, Talawanda Monroe$^2$, 
Sowgat Muzahid$^{25}$,
Benjamin Oppenheimer$^{4}$, Molly Peeples$^{2,26}$, C\'eline P\'eroux$^{16}$, 
Patrick Petitjean$^{27}$, Andreea Petric$^2$, 
Max Pettini$^{28}$, Zhijie Qu$^{29}$, Kate Rowlands$^{1,26}$, 
Ravi Sankrit$^2$, Debopam Som$^2$, 
Raghunathan Srianand$^{25}$, Nicolas Tejos$^{30}$, Jason Tumlinson$^{2}$, 
Bart Wakker$^{31}$, Jessica Werk$^{32}$}

\vspace{0.1cm}
\noindent\small 
1: STScI/ESA, 2: STScI, 3: University of Potsdam, 
4: University of Colorado, 5: Presbyterian College, 
6: University of Alabama, 7: Virginia Tech, 
8: AIP Postdam, 9: Texas Christian University, 
10: NC State, 11: Arizona State, 
12: New Mexico State, 13: Pennsylvania State, 
14: University of Chicago, 15: Durham University,
16: ESO, 17: University of Edinburgh, 
18: Tel Aviv University, 19: Caltech, 
20: University of Michigan, 21: Swinburne University,
22: Indian Institute of Technology, Tirupati, 23: University of South Carolina,
24: University of Notre Dame, 25: IUCAA, India,
26: Johns Hopkins University, 27: IAP, Paris, 
28: University of Cambridge, 29: Tsinghua University, 
30: PUCV, Chile, 31: Eureka Scientific,
32: University of Washington  
\normalsize

\begin{tcolorbox}[width=\linewidth, sharp corners=all, colback=white!95!black]
{\bf Executive Summary}: Hubble is still in prime observing condition for making 
transformative discoveries in UV astronomy.
In this white paper we describe the science case for a deep (S/N$\gtrsim$30) 
UV spectroscopic survey 
with HST/COS targeting $\approx$20 QSOs at $0.5\!<\!z\!<\!1.5$ at 
good resolution ($\approx$20\kms). 
This survey would capitalize on our current UV capability, 
produce a legacy dataset enabling community science in  many areas of 
galactic and extragalactic research,
and pioneer a path for future UV science with the Habitable Worlds Observatory. 
Such high-S/N spectra are largely missing from the MAST archives, and would be analogous to 
the deep Hubble imaging fields (HDF, UDF, Frontier Fields) that have been enormously 
successful and far-reaching in their science impact. 
This legacy dataset would enable 
frontier science programs in several areas: 

\begin{itemize}
\vspace{-0.3cm}    
\item {\bf CGM/IGM:} S/N$\gtrsim$30 spectra would provide unparalleled sensitivity 
to diffuse gas, covering a wide range of UV metal lines 
and reaching very low \ion{H}{1} column densities of log\,$N\approx12.6$
and low metallicities [Z/H]$\approx\!-2$.
This will enable precision studies of the chemical abundances, ionization, temperature, and
baryon and metal budgets of the CGM and IGM.

\vspace{-0.3cm}    
\item {\bf Milky Way \& Local Group:} Each spectrum would probe the entire quasar line-of-sight and trace foreground gas in the Local Group with extreme sensitivity, including in the Galactic ISM and halo, high-velocity clouds, and gas streams from satellite mergers.

\vspace{-0.3cm}    
\item {\bf AGN:} The redshift range $0.5\!<\!z\!<\!1.5$ will explore the properties of active galactic 
nuclei (AGN) in the extreme ultraviolet (EUV) wavelength range, covering continuum-generation
mechanisms and diagnostics of gas in accretion-disk outflows.
\end{itemize}
\end{tcolorbox}

\clearpage
{\bf Introduction:} A prime science driver for the Habitable Worlds Observatory 
is dissecting the cosmic web by probing hundreds of UV sightlines through 
galactic ecosystems, IGM filaments, and voids. This is necessary to
complete the picture of cosmic baryonic evolution --- how gas aggregates, accretes
onto galaxies, and circulates back, enriched with newly-synthesized metals from star formation ---
a priority area in the Astro2020 Decadal Survey.
The patterns of density, temperature, and metallicity are all crucial tests in theories of galaxy 
formation. 
Future X-ray telescopes (Lynx, NewAthena) plan to detect the hotter phases 
of this gas as a key scientific goal, 
and the upcoming UVEX mission will have some spectroscopic capability at low resolution,
but Hubble UV observations can already probe this multi-phase gas, over a large range 
of redshift, at good resolution and S/N. {\bf No other facility until Habitable Worlds has 
a similar ability to probe and characterize cosmic ecosystems as Hubble.}

\vspace{-0.2cm}
In the 17 years of COS operations, several large surveys have vastly improved our knowledge of the low$-z$ CGM ($z\!<\!0.5$). COS-Halos \citep{Tumlinson2013}, COS-Dwarfs \citep{Bordoloi2014}, 
CasBAH \citep{Burchett2019}, COS-GASS \citep{Borthakur2015}, CUBS \citep{Chen2020, Zahedy2021}, 
and other surveys have observed $\approx$100-150 AGN sightlines, but typically only to depths of 
S/N$\approx$10—20 at most. 
Datasets that probe Lyman Limit Systems \citep{Shull2017, Wotta2019} and the IGM 
\citep{Danforth2016, Rossenwasser2025} have similar ranges of 
S/N ratios. While these surveys have been transformational, none of them pushed the observational 
limits of COS, which can reach S/N$\approx$25--30 without special flat fields. 
In this white paper we argue this is a missed opportunity, and that an investment of
HST observing time in high-S/N UV spectroscopy of QSOs at $0.5\!<\!z\!<\!1.5$ would 
yield a significant scientific return in many areas of astronomy and astrophysics. 
By targeting this redshift range, 
we will complement ground-based datasets from Keck, VLT, and Magellan, whose archives contain many high-S/N QSO spectra at $z\!>\!2$. The advantages of the $0.5\!<\!z\!<\!1.5$ range are many: uncluttered sightlines, galaxy surveys down to fainter galaxy magnitudes, the ability to trace galaxy evolution over most of the age of the universe across the peak of cosmic star-formation activity, and access to many key metal line transitions below Ly$\alpha$ emission, which are blended by the Ly$\alpha$ forest at higher $z$.

We now describe three science areas that would significantly benefit from an HST/COS database of 
$\approx$20 QSOs at $0.5\!<\!z\!<\!1.5$ observed at S/N$\gtrsim$30, covering the range $\approx$1150--1700\AA. 
\begin{figure}[!ht]
\begin{center} 
\includegraphics[width=0.95\linewidth]{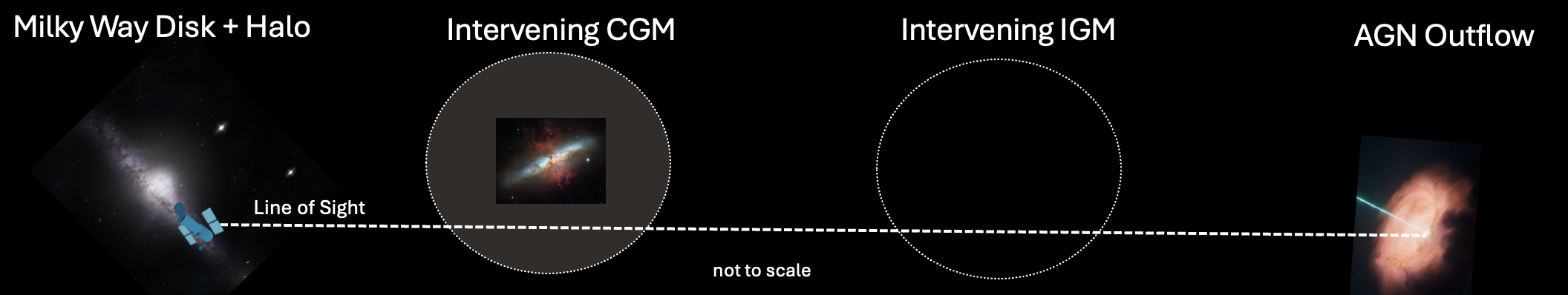}
\includegraphics[width=0.28\linewidth]{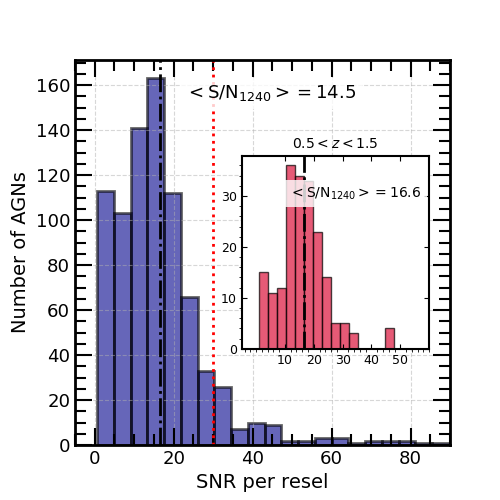}
\includegraphics[width=0.67\linewidth]{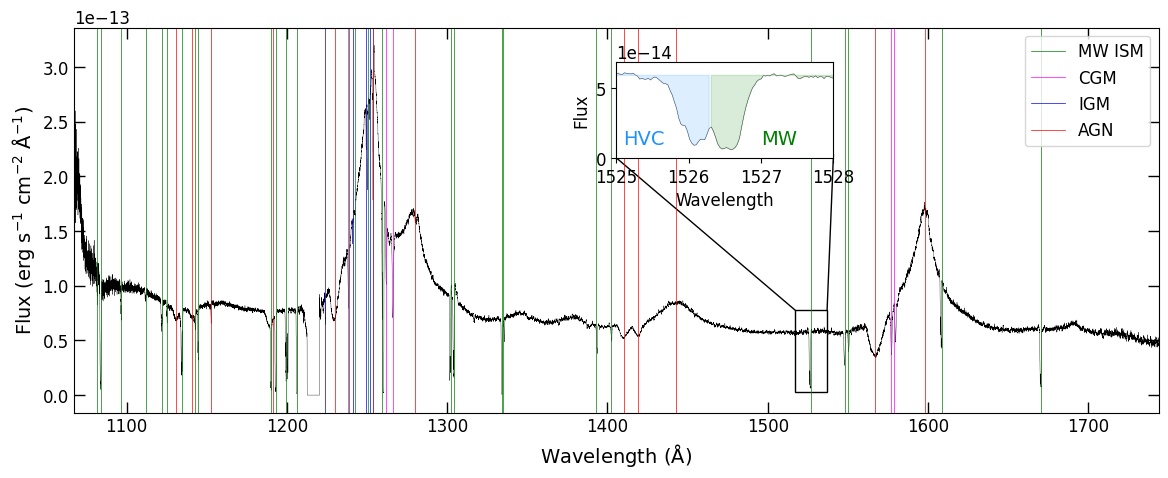}
\caption{{\bf Top:} Schematic representation of the four astrophysical environments probed by each quasar spectrum. 
{\bf Lower Left:} Distribution of S/N ratios in all (1040) COS/FUV G130M/G160M AGN spectra in the MAST archive as of December 2025.
The mean S/N is 14.5 per resolution element, and there
is a shortage of high S/N spectra ($\gtrsim$30),
particularly for targets at $0.5\!<\!z\!<\!1.5$ (inset panel).
{\bf Lower Right:} The highest-S/N COS AGN spectrum currently available, 
of the Seyfert Galaxy Mrk 817 
\citep{fox2023},
formed by co-adding 165 individual exposures. 
Absorption lines from each of the four environments are shown with color-coded vertical lines,
demonstrating the very high information density of the spectra.
Each MW line has several absorption components including high-velocity clouds (HVCs), as shown in the inset panel.
A uniform set of high-S/N HST/COS AGN spectra at $0.5\!<\!z\!<\!1.5$ would yield large scientific payoffs across all four scientific areas shown and would represent ``deep fields'' for spectroscopy.} 
\label{figure:snr}
\end{center}
\end{figure}

\vspace{-0.2cm}
{\bf CGM/IGM Science:} 
A deep survey 
would provide unparalleled
sensitivity to diffuse CGM and IGM gas, reaching very low \ion{H}{1} column 
densities of log\,($N$/cm$^{-2})\approx12.6$ and a wide range of UV metal lines across 
many ionization states. This will enable studies of:
(1) the chemical abundances, ionization state, and baryon and metal budgets of the CGM and IGM with unprecedented precision;
(2) the statistics of the low-$z$ Lyman-$\alpha$ forest, which encodes information on heating and photoionization by ionizing radiation \citep{Danforth2016};
(3) observations of the rest-frame EUV lines of the high-ionization Li-like 
ions \ion{Ne}{8} $\lambda\lambda$ 770,780, \ion{Mg}{10} $\lambda\lambda$ 609,624, 
and \ion{Si}{12} $\lambda\lambda$ 499,520, which  
at $0.5\!<\!z\!<\!1.5$ are observable in the far UV \citep{Burchett2019}.
These lines probe warm-hot gas (log $T$=5.5—6.4) in the IGM and CGM (the WHIM), which
is predicted to contain a significant fraction of cosmic baryons; 
(4) broad Lyman-alpha (BLA) absorbers \citep{Richter2006},
which trace hot gas directly without needing a metallicity correction, 
but which need high S/N for detection;
(5) important IGM coolants including \ion{O}{3}, \ion{O}{4}, and \ion{O}{5},
whose rest-frame EUV lines are redshifted into the FUV for our proposed redshift range. 
These lines can be used to quantify the IGM cooling rate.

\vspace{-0.2cm}
{\bf MW and Local Group Science} would benefit substantially from a high-S/N legacy 
database of QSO spectroscopy. These spectra would allow the chemical abundances and 
physical conditions of Galactic ISM and halo clouds to be derived in exquisite detail, 
because high S/N data cover more metal lines over a wider range of chemical elements 
and ionization states. 
As an example, an extremely high S/N ($>$200 per resolution element near 1500 \AA) coadded 
spectrum of the QSO Mrk 817 taken from the AGN-STORM2 program \citep{Kara2021} enabled a 
chemical abundance analysis of MW halo gas at unprecedented precision \citep{fox2023}, 
revealing the first evidence for dust depletion in high-velocity cloud Complex C.
Furthermore, the deep QSO spectra would cover portions of the
H$_2$ Lyman-Warner bands, useful for the study
of the molecular ISM and star formation in low-metallicity regions.

\vspace{-0.2cm}
{\bf AGN Science} would also benefit substantially. The accretion-disk continuum
peaks in the EUV, and its shape is determined
by the competing effects of its own warm, Comptonizing atmosphere and 
irradiation by hard X-rays from the innermost region surrounding the 
black hole \citep{Kubota2018}.
This survey would provide detailed spectra that have only been explored using 
composites of many lower S/N observations \citep{Zheng1997, Telfer2002, Shull2012, Stevans2014}.
It would also provide diagnostics of outflowing winds either driven off the accretion disk 
or from surrounding starburst regions. The rest-frame EUV contains a plethora of 
density-sensitive, high-ionization lines 
\citep[including metastable and excited-state lines;][]{Arav2020} such as
\ion{O}{4}, \ion{O}{5}, \ion{Ar}{6}, \ion{Ne}{5}, \ion{Ne}{6}, 
\ion{Ca}{4}, \ion{Ca}{5}, \ion{Ca}{6}, \ion{Ca}{7}, \ion{Ca}{8},
\ion{Ne}{8}, \ion{Mg}{10}, and \ion{Si}{12}.
Diagnostics from these lines can distinguish between a high-ionization component of AGN outflows typically
associated with X-ray warm absorbers and starburst winds. 
The proposed spectra would also cover H$_2$ and excited lines of \ion{Si}{2}.

\vspace{-0.2cm}
{\bf The Need for High S/N}: The vast majority of COS AGN spectra in MAST have 
low-to-moderate S/N.
Figure~\ref{figure:snr} (bottom-left panel) shows the S/N distribution of all (1040) COS/FUV G130M/G160M spectra of AGN 
and QSOs in the MAST archive as of December 2025 
(very similar to HSLA Data Release 2).
The mean S/N is 14.5, and the distribution falls off rapidly above S/N$\approx$30, 
particularly for AGN at $0.5\!<\!z\!<\!1.5$,
preventing the kind of CGM/IGM science that requires high sensitivity. 
High-S/N data have several key advantages. 
They provide sensitivity to weak absorption, vastly increasing the number of UV metal line diagnostics that can be detected in the CGM and IGM and enabling studies of the detailed chemical composition and dust depletion in the gas, and probing \ion{H}{1} 
down to under-dense cosmic regions (log\,$N$(\hi)$\lesssim$13).
When coupled with the medium-to-high spectral resolution ($\approx$15--20\kms) of COS G130M/G160M observations, high S/N data also provide strong discriminating power for line profile analysis,
revealing multiphase gas based on mismatches in the absorption line profiles of low- and high-ionization species, in terms of line centroids, line width, and asymmetry. 
The upcoming UVEX mission will not have high-resolution spectroscopic capability, so HST is the only mission in the current or near-term future capable of high-S/N high-resolution UV spectroscopy. 

\vspace{-0.2cm}  
{\bf Target Selection:} The pool of QSO targets at $0.5\!<\!z\!<\!1.5$ is large, with fluxes ranging from $2\times10^{-15}$ to $1\times10^{-14}$ erg\,s$^{-1}$\,cm$^{-2}$\,\AA$^{-1}$. Cross-matching GALEX FUV/NUV photometry with 
QSO catalogs such as the UV-bright QSO Survey \citep[UVQS;][]{Monroe2016} 
and the Plane Quasar Survey \citep{Werk2024} yields an extensive list of candidates. 

\vspace{-0.2cm}
{\bf FUV Orbit Requirements:} We propose setting the exposure requirements to be 
S/N$\gtrsim$30 per resolution element everywhere between 1275 and 1535 \AA, 
lowering to $\approx$20 at longer and shorter wavelengths and rising higher 
at the peak throughputs for each grating. 
Although current FP-POS restrictions limit the S/N of individual COS FUV G130M exposures to 
$\approx$20--30 depending on setting,
multiple exposures can be co-added to reach S/N=30 or above
provided that fixed-pattern noise is accounted for,
which enables absorption-line measurements down to observed-frame 
equivalent widths of $\approx$10--15 m\AA. 
To derive anticipated FUV orbit requirements, we use the COS ETC v34.1.1. 
For a flux of $1\times10^{-14}$\,erg\,s$^{-1}$\,cm$^{-2}$\,\AA$^{-1}$, 
S/N=30 per 6-pixel resel at 1275\AA\ using G130M 
requires 18730 sec. A similar S/N at 1535\AA\ using G160M requires 40650 sec. With typical 
overheads, this is $\approx$10 and 20 orbits, respectively. In this flux range, 
the dark rate is low enough that exposure times scale linearly with flux, so the faintest practical targets
would require $\approx$20 and 40 orbits, respectively. 
A sample of $\approx$20 QSOs spanning this brightness range would require $\approx$600--1000 orbits.
The trade-offs between sample size, S/N ratio, and target redshift could be explored by a detailed parameter study.
Such a study is beyond the scope of this white paper, whose main point is to highlight the broad range of cutting-edge science that a database of high-S/N COS spectra of high-$z$ AGN would enable in several areas of astronomy and astrophysics.

\vspace{-0.2cm}
{\bf Supporting Multi-Wavelength Observations:} 
This survey would likely spur a multitude of supporting
observations, engaging a wide fraction of the astronomical community:
\begin{itemize}
\vspace{-0.5cm}  
\item 
Existing and future ground-based optical, radio, and mm surveys can be used to identify galaxies corresponding to detected absorption lines 
\citep[e.g. the CUBS sample;][]{Chen2020}.
\vspace{-0.3cm}  
\item Complementary HST or JWST imaging and grism spectroscopy could measure the
absorbing galaxies' orientation to reveal the CGM's shape and dynamics in different ionization states.
\vspace{-0.3cm}  
\item Deep X-ray spectroscopy can probe the multiphase components of any AGN outflows.
\vspace{-0.3cm}  
\item Deep X-ray imaging observations could probe X-ray scattering halos around
the AGN, giving an independent measure of gray dust along the sightline and better constraining the IGM state.
\end{itemize}

\vspace{-0.4cm}
\small
\bibliographystyle{aasjournal}
\bibliography{references}

@ARTICLE{fox2023,
       author = {{Fox}, Andrew J. and {Cashman}, Frances H. and {Kriss}, Gerard A. and {et al.}},
        title = "{Detection of Dust in High-velocity Cloud Complex C-Enriched Gas Accreting onto the Milky Way}",
      journal = {\apjl},
         year = 2023,
        month = apr,
       volume = {946},
       number = {2},
          eid = {L48},
        pages = {L48},
          doi = {10.3847/2041-8213/acc640},
archivePrefix = {arXiv},
       eprint = {2303.12577},
 primaryClass = {astro-ph.GA},
       adsurl = {https://ui.adsabs.harvard.edu/abs/2023ApJ...946L..48F}
}

@ARTICLE{Burchett2019,
       author = {{Burchett}, Joseph N. and {Tripp}, Todd M. and {Prochaska}, J. Xavier and {et al.}},
        title = "{The COS Absorption Survey of Baryon Harbors (CASBaH): Warm-Hot Circumgalactic Gas Reservoirs Traced by Ne VIII Absorption}",
      journal = {\apjl},
         year = 2019,
        month = jun,
       volume = {877},
       number = {2},
          eid = {L20},
        pages = {L20},
          doi = {10.3847/2041-8213/ab1f7f},
archivePrefix = {arXiv},
       eprint = {1810.06560},
 primaryClass = {astro-ph.GA},
       adsurl = {https://ui.adsabs.harvard.edu/abs/2019ApJ...877L..20B}
}

@ARTICLE{Bordoloi2014,
       author = {{Bordoloi}, Rongmon and {Tumlinson}, Jason and {Werk}, Jessica K. and {et al.}},
        title = "{The COS-Dwarfs Survey: The Carbon Reservoir around Sub-L* Galaxies}",
      journal = {\apj},
         year = 2014,
        month = dec,
       volume = {796},
       number = {2},
          eid = {136},
        pages = {136},
          doi = {10.1088/0004-637X/796/2/136},
archivePrefix = {arXiv},
       eprint = {1406.0509},
 primaryClass = {astro-ph.GA},
       adsurl = {https://ui.adsabs.harvard.edu/abs/2014ApJ...796..136B}
}

@ARTICLE{Borthakur2015,
       author = {{Borthakur}, Sanchayeeta and {Heckman}, Timothy and {Tumlinson}, Jason and {et al.}},
        title = "{Connection between the Circumgalactic Medium and the Interstellar Medium of Galaxies: Results from the COS-GASS Survey}",
      journal = {\apj},
         year = 2015,
        month = nov,
       volume = {813},
       number = {1},
          eid = {46},
        pages = {46},
          doi = {10.1088/0004-637X/813/1/46},
archivePrefix = {arXiv},
       eprint = {1504.01392},
 primaryClass = {astro-ph.GA},
       adsurl = {https://ui.adsabs.harvard.edu/abs/2015ApJ...813...46B}
}

@ARTICLE{Arav2020,
       author = {{Arav}, Nahum and {Xu}, Xinfeng and {Miller}, Timothy and {Kriss}, Gerard A. and {Plesha}, Rachel},
        title = "{HST/COS Observations of Quasar Outflows in the 500-1050 {\r{A}} Rest Frame. I. The Most Energetic Outflows in the Universe and Other Discoveries}",
      journal = {\apjs},
         year = 2020,
        month = apr,
       volume = {247},
       number = {2},
          eid = {37},
        pages = {37},
          doi = {10.3847/1538-4365/ab66af},
archivePrefix = {arXiv},
       eprint = {2003.08688},
 primaryClass = {astro-ph.GA},
       adsurl = {https://ui.adsabs.harvard.edu/abs/2020ApJS..247...37A}
}

@ARTICLE{Tumlinson2013,
       author = {{Tumlinson}, Jason and {Thom}, Christopher and {Werk}, Jessica K. and 
       {et al.}},
        title = "{The COS-Halos Survey: Rationale, Design, and a Census of Circumgalactic Neutral Hydrogen}",
      journal = {\apj},
         year = 2013,
        month = nov,
       volume = {777},
       number = {1},
          eid = {59},
        pages = {59},
          doi = {10.1088/0004-637X/777/1/59},
archivePrefix = {arXiv},
       eprint = {1309.6317},
 primaryClass = {astro-ph.CO},
       adsurl = {https://ui.adsabs.harvard.edu/abs/2013ApJ...777...59T}
}

@ARTICLE{Monroe2016,
       author = {{Monroe}, TalaWanda R. and {Prochaska}, J. Xavier and {Tejos}, Nicolas and {et al.}},
        title = "{The UV-bright Quasar Survey (UVQS): DR1}",
      journal = {\aj},
         year = 2016,
        month = jul,
       volume = {152},
       number = {1},
          eid = {25},
        pages = {25},
          doi = {10.3847/0004-6256/152/1/25},
archivePrefix = {arXiv},
       eprint = {1602.06255},
 primaryClass = {astro-ph.GA},
       adsurl = {https://ui.adsabs.harvard.edu/abs/2016AJ....152...25M}
}

@ARTICLE{Chen2020,
       author = {{Chen}, Hsiao-Wen and {Zahedy}, Fakhri S. and {Boettcher}, Erin and {et al.}},
        title = "{The Cosmic Ultraviolet Baryon Survey (CUBS) - I. Overview and the diverse environments of Lyman limit systems at z < 1}",
      journal = {\mnras},
         year = 2020,
        month = sep,
       volume = {497},
       number = {1},
        pages = {498-520},
          doi = {10.1093/mnras/staa1773},
archivePrefix = {arXiv},
       eprint = {2005.02408},
 primaryClass = {astro-ph.GA},
       adsurl = {https://ui.adsabs.harvard.edu/abs/2020MNRAS.497..498C}
}

@ARTICLE{Werk2024,
       author = {{Werk}, Jessica and {Tchernyshyov}, Kirill and {Bish}, Hannah and {et al.}},
        title = "{The Plane Quasar Survey: First Data Release}",
      journal = {\apjs},
         year = 2024,
        month = aug,
       volume = {273},
       number = {2},
          eid = {21},
        pages = {21},
          doi = {10.3847/1538-4365/ad58df},
archivePrefix = {arXiv},
       eprint = {2403.12266},
 primaryClass = {astro-ph.GA},
       adsurl = {https://ui.adsabs.harvard.edu/abs/2024ApJS..273...21W}
}

@ARTICLE{Kara2021,
       author = {{Kara}, Erin and {Mehdipour}, Missagh and {Kriss}, Gerard A. and {et al.}},
        title = "{AGN STORM 2. I. First results: A Change in the Weather of Mrk 817}",
      journal = {\apj},
         year = 2021,
        month = dec,
       volume = {922},
       number = {2},
          eid = {151},
        pages = {151},
          doi = {10.3847/1538-4357/ac2159},
archivePrefix = {arXiv},
       eprint = {2105.05840},
 primaryClass = {astro-ph.HE},
       adsurl = {https://ui.adsabs.harvard.edu/abs/2021ApJ...922..151K}
}

@ARTICLE{Danforth2016,
       author = {{Danforth}, Charles W. and {Keeney}, Brian A. and {Tilton}, Evan M. and {et al.}},
        title = "{An HST/COS Survey of the Low-redshift Intergalactic Medium. I. Survey, Methodology, and Overall Results}",
      journal = {\apj},
         year = 2016,
        month = feb,
       volume = {817},
       number = {2},
          eid = {111},
        pages = {111},
          doi = {10.3847/0004-637X/817/2/111},
archivePrefix = {arXiv},
       eprint = {1402.2655},
 primaryClass = {astro-ph.CO},
       adsurl = {https://ui.adsabs.harvard.edu/abs/2016ApJ...817..111D}
}

@ARTICLE{Richter2006,
       author = {{Richter}, P. and {Savage}, B.~D. and {Sembach}, K.~R. and {Tripp}, T.~M.},
        title = "{Tracing baryons in the warm-hot intergalactic medium with broad Ly {\ensuremath{\alpha}} absorption}",
      journal = {\aap},
         year = 2006,
        month = jan,
       volume = {445},
       number = {3},
        pages = {827-842},
          doi = {10.1051/0004-6361:20053636},
archivePrefix = {arXiv},
       eprint = {astro-ph/0509539},
 primaryClass = {astro-ph},
       adsurl = {https://ui.adsabs.harvard.edu/abs/2006A&A...445..827R}
}

@ARTICLE{Zahedy2021,
       author = {{Zahedy}, Fakhri S. and {Chen}, Hsiao-Wen and {Cooper}, Thomas M. and {et al.}},
        title = "{The Cosmic Ultraviolet Baryon Survey (CUBS) - III. Physical properties and elemental abundances of Lyman-limit systems at z < 1}",
      journal = {\mnras},
         year = 2021,
        month = sep,
       volume = {506},
       number = {1},
        pages = {877-902},
          doi = {10.1093/mnras/stab1661},
archivePrefix = {arXiv},
       eprint = {2106.04608},
 primaryClass = {astro-ph.GA},
       adsurl = {https://ui.adsabs.harvard.edu/abs/2021MNRAS.506..877Z}
}

@ARTICLE{Kubota2018,
       author = {{Kubota}, Aya and {Done}, Chris},
        title = "{A physical model of the broad-band continuum of AGN and its implications for the UV/X relation and optical variability}",
      journal = {\mnras},
         year = 2018,
        month = oct,
       volume = {480},
       number = {1},
        pages = {1247-1262},
          doi = {10.1093/mnras/sty1890},
archivePrefix = {arXiv},
       eprint = {1804.00171},
 primaryClass = {astro-ph.HE},
       adsurl = {https://ui.adsabs.harvard.edu/abs/2018MNRAS.480.1247K}
}

@ARTICLE{Rossenwasser2025,
       author = {{Rosenwasser}, Benjamin E. and {Wakker}, Bart P. and {Savage}, Blair D.},
        title = "{A Classification Scheme for Quasar Absorption Lines}",
      journal = {\apj},
         year = 2025,
        month = may,
       volume = {985},
       number = {1},
          eid = {31},
        pages = {31},
          doi = {10.3847/1538-4357/adc4ea},
       adsurl = {https://ui.adsabs.harvard.edu/abs/2025ApJ...985...31R}
}

@ARTICLE{Zheng1997,
       author = {{Zheng}, Wei and {Kriss}, Gerard A. and {Telfer}, Randal C. and {Grimes}, John P. and {Davidsen}, Arthur F.},
        title = "{A Composite HST Spectrum of Quasars}",
      journal = {\apj},
         year = 1997,
        month = feb,
       volume = {475},
       number = {2},
        pages = {469-478},
          doi = {10.1086/303560},
archivePrefix = {arXiv},
       eprint = {astro-ph/9608198},
 primaryClass = {astro-ph},
       adsurl = {https://ui.adsabs.harvard.edu/abs/1997ApJ...475..469Z}
}

@ARTICLE{Telfer2002,
       author = {{Telfer}, Randal C. and {Zheng}, Wei and {Kriss}, Gerard A. and {Davidsen}, Arthur F.},
        title = "{The Rest-Frame Extreme-Ultraviolet Spectral Properties of Quasi-stellar Objects}",
      journal = {\apj},
         year = 2002,
        month = feb,
       volume = {565},
       number = {2},
        pages = {773-785},
          doi = {10.1086/324689},
archivePrefix = {arXiv},
       eprint = {astro-ph/0109531},
 primaryClass = {astro-ph},
       adsurl = {https://ui.adsabs.harvard.edu/abs/2002ApJ...565..773T}
}

@ARTICLE{Shull2012,
       author = {{Shull}, J. Michael and {Stevans}, Matthew and {Danforth}, Charles W.},
        title = "{HST-COS Observations of AGNs. I. Ultraviolet Composite Spectra of the Ionizing Continuum and Emission Lines}",
      journal = {\apj},
         year = 2012,
        month = jun,
       volume = {752},
       number = {2},
          eid = {162},
        pages = {162},
          doi = {10.1088/0004-637X/752/2/162},
archivePrefix = {arXiv},
       eprint = {1204.3908},
 primaryClass = {astro-ph.CO},
       adsurl = {https://ui.adsabs.harvard.edu/abs/2012ApJ...752..162S}
}

@ARTICLE{Wotta2019,
       author = {{Wotta}, Christopher B. and {Lehner}, Nicolas and {Howk}, J. Christopher and {O'Meara}, John M. and {Oppenheimer}, Benjamin D. and {Cooksey}, Kathy L.},
        title = "{The COS CGM Compendium. II. Metallicities of the Partial and Lyman Limit Systems at z {\ensuremath{\lesssim}} 1}",
      journal = {\apj},
         year = 2019,
        month = feb,
       volume = {872},
       number = {1},
          eid = {81},
        pages = {81},
          doi = {10.3847/1538-4357/aafb74},
archivePrefix = {arXiv},
       eprint = {1811.10654},
 primaryClass = {astro-ph.GA},
       adsurl = {https://ui.adsabs.harvard.edu/abs/2019ApJ...872...81W}
}

@ARTICLE{Shull2017,
       author = {{Shull}, J. Michael and {Danforth}, Charles W. and {Tilton}, Evan M. and {Moloney}, Joshua and {Stevans}, Matthew L.},
        title = "{An Ultraviolet Survey of Low-redshift Partial Lyman-limit Systems with the HST Cosmic Origins Spectrograph}",
      journal = {\apj},
         year = 2017,
        month = nov,
       volume = {849},
       number = {2},
          eid = {106},
        pages = {106},
          doi = {10.3847/1538-4357/aa9229},
archivePrefix = {arXiv},
       eprint = {1710.03232},
 primaryClass = {astro-ph.GA},
       adsurl = {https://ui.adsabs.harvard.edu/abs/2017ApJ...849..106S}
}

@ARTICLE{Stevans2014,
       author = {{Stevans}, Matthew L. and {Shull}, J. Michael and {Danforth}, Charles W. and {Tilton}, Evan M.},
        title = "{HST-COS Observations of AGNs. II. Extended Survey of Ultraviolet Composite Spectra from 159 Active Galactic Nuclei}",
      journal = {\apj},
         year = 2014,
        month = oct,
       volume = {794},
       number = {1},
          eid = {75},
        pages = {75},
          doi = {10.1088/0004-637X/794/1/75},
archivePrefix = {arXiv},
       eprint = {1408.5900},
 primaryClass = {astro-ph.GA},
       adsurl = {https://ui.adsabs.harvard.edu/abs/2014ApJ...794...75S}
}
\normalsize

\end{document}